# Title: Battery Diagnostics with Sensitive Magnetometry

**Authors:**
Yinan Hu[1,2,#], Geoffrey Z. Iwata[1,2,#*], Mohaddese Mohammadi[3], Emilia V. Silletta[3], Arne Wickenbrock[1,2], John W. Blanchard[2], Dmitry Budker[1,2,4,5], Alexej Jerschow[3*]

**Affiliations:**
[1]Johannes Gutenberg-Universität Mainz, 55099 Mainz, Germany
[2]Helmholtz Institute Mainz, 55128 Mainz, Germany
[3]Department of Chemistry, New York University, New York, NY 10003
[4]Department of Physics, University of California Berkeley, Berkeley, CA 94720-7300, USA
[5]Nuclear Science Division, E. O. Lawrence National Laboratory, Berkeley, CA 94720, USA

# these authors contributed equally (joint first authors)

**\*** Corresponding authors:
Geoffrey Iwata, iwata@uni-mainz.de
Alexej Jerschow, alexej.jerschow@nyu.edu

**Abstract**
The ever-increasing demand for high-capacity rechargeable batteries highlights the need for sensitive and accurate diagnostic technology for determining the state of a cell, for identifying and localizing defects, or for sensing capacity loss mechanisms. Here, we demonstrate the use of atomic magnetometry to map the weak induced magnetic fields around a Li-ion battery cell as a function of state of charge and upon introducing mechanical defects. These measurements provide maps of the magnetic susceptibility of the cell, which follow trends characteristic for the battery materials under study upon discharge. In addition, the measurements reveal hitherto unknown long time-scale transient internal current effects, which were particularly pronounced in the overdischarged regime. The diagnostic power of this technique is promising for the assessment of cells in research, quality control, or during operation, and could help uncover details of charge storage and failure processes in cells.

**Main text**

Rechargeable batteries lie at the heart of technological developments, enabling the use of renewable energy sources, powering electric vehicles, cell phones, and other portable electronics[1]. Key to the success of many rechargeable battery-powered developments is the reliable fabrication and deployment of battery cells with sufficient capacities and lifetimes. One of the biggest challenges in this process, however, is to determine the quality of cells by nondestructive measurements. Currently-used techniques which can give *in situ/ operando* information include synchrotron-based scanning transmission X-ray microscopy[2], X-ray micro-diffraction[3], and neutron diffraction and Raman spectroscopy[4]. Manufacturers typically perform electrochemical testing and limited 2D X-ray scanning[5,6]. X-ray tomography is a potentially powerful tool as well for commercial cells but is generally too slow for high throughput use[5,7]. Ultrasound-based measurements are promising for characterization of cells based on changes in density and mechanical properties[8,9].

Often critical device parameters elude many such examinations. In particular, once cells are compromised by mechanical impact, overcharging, or otherwise reaching their end of life, it is particularly difficult to investigate the causes and the propagation of failure modes[10]. Destructive analysis can provide important clues, but materials can be significantly altered in the process. Furthermore, this approach precludes a study of cells over time and does not give access to markers for cell lifetimes[11].

Recently, it was demonstrated that magnetic susceptibility changes within cells could be measured non-destructively using an inside-out MRI (ioMRI) technique[12] that used the $^1$H nuclear spin resonance frequencies in water to measure the susceptibility-induced field surrounding a cell when placed in a strong magnetic field. It was shown that the changes in the magnetic susceptibility could be tracked across the charge-discharge cycle, and that these changes followed the expected trends of the lithiation state of the cathode material. This approach thus provided a single-point state-of-charge measurement and allowed for the identification of inhomogeneities or non-idealities of charge storage in electrochemical cells. While the ioMRI measurement is fast and provides high resolution, it relies on complex and costly MRI instrumentation.

In this work, we demonstrate the use of atomic magnetometers to enable measurements of the magnetic susceptibility within cells, and to report on the localized state of charge and the defects inside. Furthermore, the use of these sensors allowed the measurement of hitherto unknown transient internal currents following cell discharge. These currents were particularly pronounced in the region of overdischarge.

**Measurement setup**
Magnetic susceptibility measurements involve placing a sample in a magnetic field and measuring the smaller induced magnetic fields. At the same time, the magnetic field sensors, which typically have a limited dynamic range, need to be located in a small magnetic field (<50 nT) in order to detect the minute fields (<100 pT) induced by the cell.

The strategy employed in this work was to use a long flat solenoid to apply a magnetic field to the cell, as shown in Fig 1. The solenoid pierced a magnetically shielded region produced by concentric cylinders of mu-metal[13–15]. In this arrangement, a negligible magnetic field is produced outside of the solenoid. The magnetic field sensors are placed in this region of negligible field within the magnetic shield and operate within their dynamic range (Figure 1a). The induced magnetic field of a battery cell located inside the solenoid, however, is communicated to the sensor region without impediment. In addition to reducing environmental magnetic fields, the magnetic shield arrangement (Figure 1b) also ensures that the magnetic flux lines emerging from the ends of the solenoid connect outside of the shielded region.

Atomic magnetometers were selected as the field sensors for this study because they offer the highest sensitivity, and simultaneous multi-axis measurements. These sensors provide a measure of the magnetic field via optical detection of the atomic electron resonance frequency shifts[16]. Recently, atomic magnetometers have become commercially available in miniaturized designs[17].

To obtain a magnetic field map of a battery, the cell is moved via a "conveyor belt" within the solenoid (Figure 1b). The belt moves the battery past the sensors for scanning along the $z$-coordinate. For subsequent scans, the cell was translated along $x$ for the next scan while it was transported back to the original $z$ position. In this way, the induced field from the battery was scanned across a rectangular area above and below the cell. Figure 1c shows representative field maps for the magnetic field components, $B_x$ and $B_z$, obtained from a single sensor above the battery. These maps, recorded with the solenoid on, are consistent with maps expected for a rectangular block with approximately uniform susceptibility (similar to a dipole field map), which represents a good first approximation of a cell. Figure 1d shows the circuit setup for switching between the charge and discharge operation of the cell. The cell was connected to this circuit via twisted wires running along the conveyor belt.

Without the solenoid field applied to the battery, the setup is sensitive to internal currents within the battery, and any remnant magnetization of the battery components. Turning on the solenoid field induces a magnetization within the cell, dependent on the local magnetic susceptibility. This magnetization produces a field that is detected by the magnetometers.

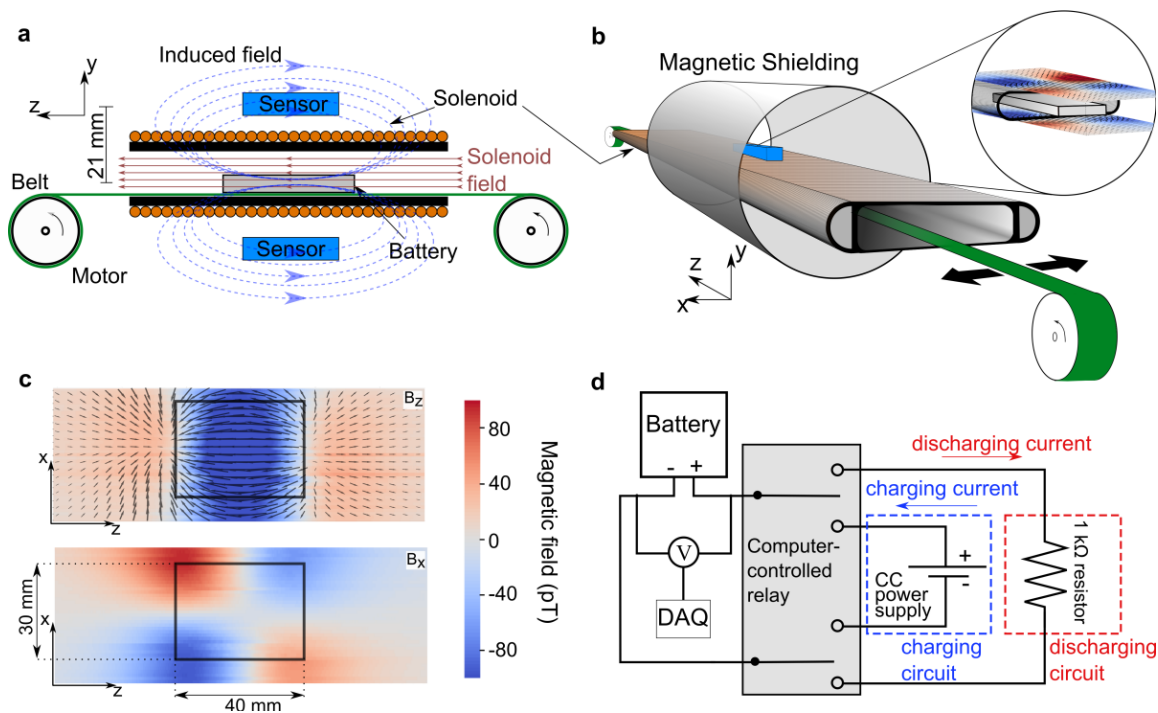

**Fig. 1: Experimental susceptometry setup. a**, Side view of the experimental arrangement. A battery is placed on a motor-driven "conveyor belt" that moves the cell through a custom-designed solenoid, which provides a constant internal magnetic field. Magnetic-field sensors are placed in the ultralow-field region, above and below the solenoid. **b**, 3D drawing of the experimental arrangement. The solenoid is shown extending beyond the Twinleaf MS-2 magnetic shielding (only the innermost shield is shown – a total of four concentric cylinders were used). The conveyor belt moves the battery back and forth along the $z$-axis, while translation stages on each end move the entire belt with 0.1 mm precision along the $x$-axis. The inset illustrates the location of the resulting two-dimensional field maps acquired by this setup. **c**, Sample-induced magnetic field map for a "healthy" battery charged to 90% capacity, obtained from a single sensor above the battery. The upper plot shows a color-map of the $z$-component of the measured field, with a vector map of the total field in the $x$-$z$ sensor plane. The lower plot shows the $x$-component color-map of the measured field. The battery outline is shown as a black rectangle. **d**, Circuit diagram for charging and discharging. The computer-controlled relay disconnects the battery cell from the charging/discharging circuits during measurements. However, the data acquisition (DAQ) unit remained connected throughout the measurement. V: Voltmeter. CC: constant current.

## Magnetic susceptibility and internal currents vs. discharge

To measure a battery's magnetic properties across a discharge cycle, the behavior of the cell voltage and the induced magnetic field were monitored during 30-minute rest periods between 30-minute discharge periods. For discharge, the battery electrodes were connected to a 1 kΩ resistor, as shown in Figure 1d. The solenoid field was on during this measurement, so the apparatus was sensitive to susceptibility-induced fields.

Figure 2a shows a series of 30-minute magnetic field measurements during the rest periods immediately after disconnecting the cell from the discharge circuit. The magnetic field values are referenced to the value at the full discharge capacity of the cells (813 mAh), extrapolated from the behavior near in the overdischarge regime. Transient fields were observed, which relaxed exponentially towards a steady-state value. The signal was fit to a single exponential function of the form, $y(x) = Ae^{-x/t_1} + c$, where $A$ is the amplitude, $t_1$ is the time constant, and $c$ is an offset representative of the long-term static value. The time constants were stable with an average of 11±4 s in the 'healthy' battery regime up to its rated capacity (600 mAh). By contrast, the time constant increased significantly to 400-750 s when discharging beyond this region. The region of overdischarge is represented by shaded regions in Figure 2. Red squares in Figure 2a indicate the fitted time constants of the transient fields. The long-term static value of the magnetic field, $c$, decreased to background levels when the solenoid field was turned off, thus must reflect the fields generated by the induced magnetic field, and specifically relate to the magnetic susceptibility of the battery components. The transient fields are present even with the solenoid turned off, indicating that they are not due to susceptibility changes in the cell. Instead, they appear to be related to internal equilibration currents.

Figure 2b shows the long-term static values of the magnetic field after each discharge period. The features at 90 mAh and 300 mAh can be attributed to laboratory noise that occurred during the measurement, as assessed by independent fluxgate field measurements outside the shield in the vicinity of the setup (marked with an asterisk in the Figure). Within the 'healthy' range (up to the rated capacity), the long-term field value changes gradually; beyond the rated discharge capacity (in the shaded region), we observe a significantly stronger effect. This behavior is in line with previous observations for these types of cells with cathode material containing a large amount of cobalt, and can be related to the change of the magnetic susceptibility over the course of discharge[12]. The supporting information provides the composition of the cell obtained via scanning electron microscopy (SEM) microanalysis measurements in Figures S1 and S2, as well as in Table S1.

Figure 2c shows the voltage measurement during the same experiment for comparison. Voltage was measured both during the discharge and during the magnetic field measurement period, when the battery was not connected to the discharge resistor. Here, too, there is a clear long-term recovery period in the shaded region, which becomes stronger with depth of discharge. We found that the time constants for this voltage measurement and those of the magnetic field measurements agreed with each other (Figure 2d) in the overdischarge region (shaded). This finding supports the notion that internal currents may be the cause of these relaxation effects. Notably, before reaching the rated discharge capacity, the time constants were more than an order of magnitude smaller than in the region of overdischarge. In this region, fitting the voltage measurements during the relaxation period requires a double exponential, while the magnetic field measurements fit reliably well with a single time constant. This finding can likely be attributed to the influence of the connected circuitry on the voltage measurements. The magnetic field measurement, by contrast, provides a contactless measure of the internal current effects and therefore enables a measurement that is decoupled from the electrical circuit, in addition to providing spatial resolution. Given the achieved measurement sensitivity (~10 pT/√Hz), transient internal currents as small as ~4 µA can be sensed with this setup. However, these measurements do not require the solenoid and can in principle be performed in a fully closed shield, in which the magnetometers achieve a sensitivity of 20 fT/√Hz, corresponding to internal currents of 8 nA.

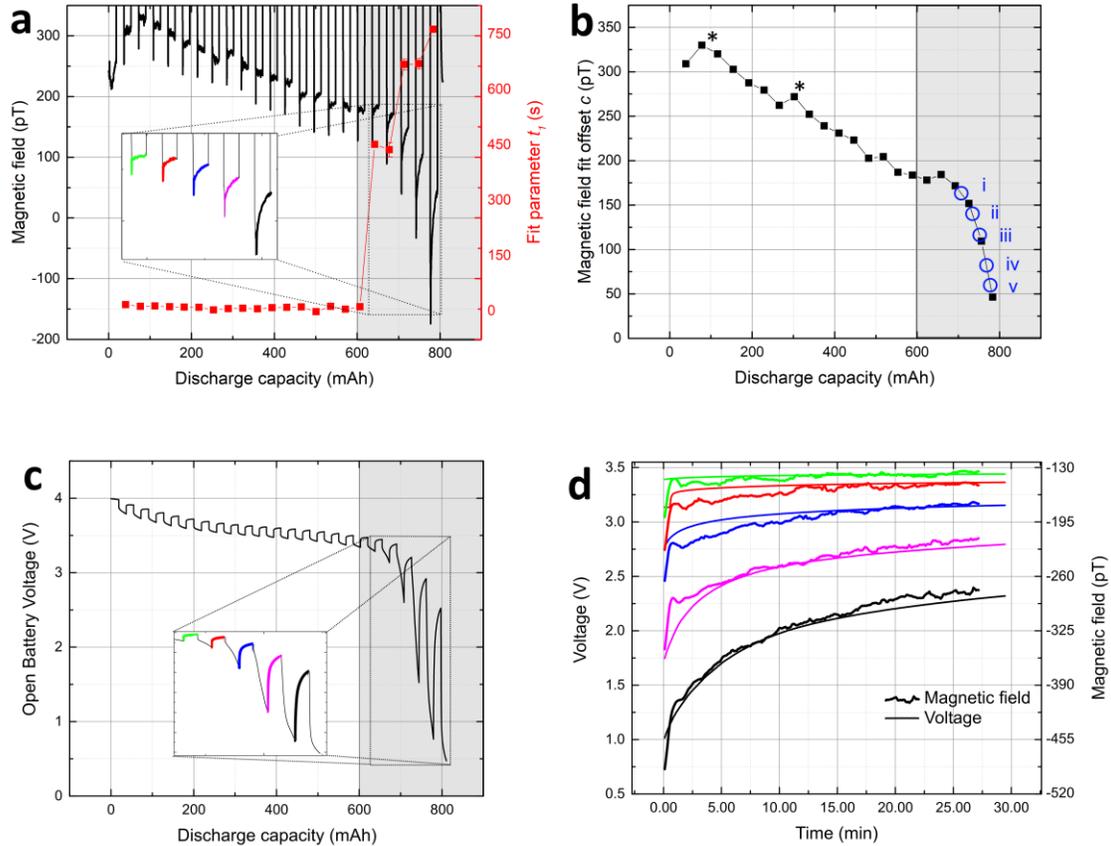

**Fig. 2: Magnetic field and voltage response of a cell after discharging.** Magnetic field measurements above the cell at a location corresponding to the upper left corner of the battery outline in Fig. 1c, and voltage measurements as a function of discharge capacity. The shaded areas indicate the overdischarge region, when the battery was measured after being discharged beyond its rated capacity (600mAh). **a,** Magnetic field, $B_x$, measured with an atomic magnetometer for periods of 30 min at the indicated discharge capacity, beginning immediately after disconnecting the discharge circuit (see Methods for details). The inset shows detail of the magnetic field decay for discharge capacities beyond the battery rating, where the equilibration time to a steady-state value increases, suggesting the presence of persistent internal currents within the cell. Red squares indicate the fitted time constants of the exponential behavior within each measurement period. Magnetic field values are referenced to a value extrapolated to the full discharge capacity of cell (813 mAh). **b**, Steady-state magnetic field values determined by fitting the 30 min measurement curves in **a** to an exponential decay. The field amplitude exhibits a gradual decrease until reaching the rated capacity and a steeper falloff in the shaded region. Roman numerals indicate the points at which the magnetic field maps of Figure 3a were acquired. Asterisks mark data points which were acquired during large environmental field fluctuations as assessed by an independent fluxgate

measurement. **c**, Voltage vs. discharge capacity. During the discharge periods, the measured voltage drops and then recovers when the battery is disconnected from the discharge circuit. The inset shows detail of the voltage signal decay for discharge capacities beyond the battery rating. **d**, Detail of the signal decay for discharge capacities beyond the battery rating (600 mAh). The colors correspond to those used in the insets in **a** and **b**.

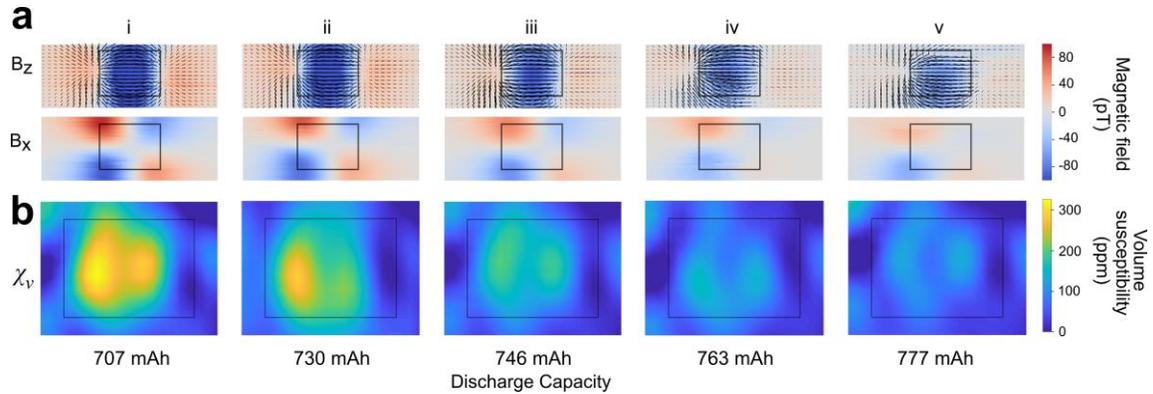

**Fig. 3: Magnetic field and corresponding susceptibility maps above a cell.** Roman numerals relate these selected snapshots to the points along the discharge curve in Figure 2b. The corresponding discharge capacity is given at the bottom. **a,** Upper panels show the field map in the *z*-direction, with an *x*-*z* field vector map overlaid. Lower panels show the field in the *x*-direction. The black rectangular outline represents the battery position. Battery leads (not indicated) are to the left of the black rectangle. **b,** 2D magnetic susceptibility maps (indicated in ppm) obtained from a regularized inversion of the measured magnetic field.

Figure 3a shows a series of 2D maps of magnetic field measurements after cell equilibration at the discharge values indicated in Figure 2b. It is again observed that the overall magnetic field decreases with discharge. Furthermore, a regularized magnetic field inversion produces magnetic susceptibility maps for each. It is observed that the magnetic susceptibility, and hence the charge is distributed non-uniformly across the cell, which could be due to the position of the battery tabs and other internal geometrical and material arrangements. Even though the overall susceptibility decreases, the main feature of two maxima in the distribution appears to persist through continued discharge. The positive values for susceptibility indicate a paramagnetic behavior of the battery cell.

**Effect of mechanical damage on induced magnetic fields**
In order to test the method for its ability to detect deviations from the susceptibility distribution, cells were subjected to physical impact and subsequently measured. A steel rod was dropped onto various points of a battery that was within 5% of full charge, and that battery was scanned in the setup to explore the effect that physical impact had on the magnetization pattern; the results are shown in Figure 4. Some of the damage resulted in visible physical deformation, but significant changes in the maps were observed even before such visible effects occurred. Throughout the tests, the cell still maintained the ability to be charged, retain energy, and power devices, despite its apparent damage. Nevertheless, the magnetic susceptibility distribution of the cell was seen to be altered.

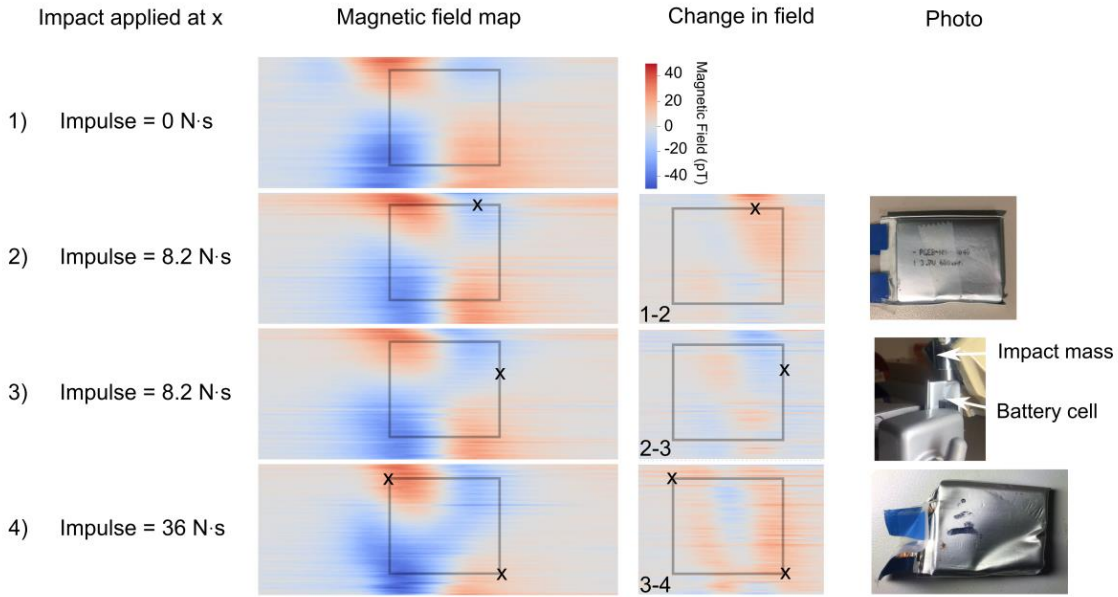

**Fig. 4: Effect of physical damage on the induced magnetic field in a battery within 5% of full charge.** A weight was dropped from a controlled height onto a specific point on the battery, indicated by the 'X'. Each measurement (1-4) was performed subsequently on the same cell. Measurement 1 shows the undamaged field map. In experiments 2-3, the battery was clamped from the flat sides while a mass was dropped from above. In the final trial, the battery was supported from the bottom and sides, resulting in two impact points, as indicated by the two 'X's.

**Alternative sensors**

Similar measurements could, in principle, be performed with other magnetometer sensor technologies. In particular, suitable candidate sensors for efficient measurement include various types of atomic magnetometers[16], magnetometers based on nitrogen-vacancy (NV) centers in diamond[18], Hall probes[19], magnetoresistive sensors[20], SQUIDs[21], and fluxgates. NV magnetometers and magnetoresistive sensors, for example, would offer a relatively high sensitivity over a large dynamic range, useful for batteries containing magnetic materials or those which exhibit large changes in susceptibility as they are charged or discharged[22]. Furthermore, the small size of these sensors could allow for a higher spatial resolution measurement of the induced magnetic field. These can be used in a sensor array to reduce the mapping time, depending on the minimal stand-off distance of the sensor. Since microwaves that are typically used in NV measurements may be undesirable in some applications, recently-developed microwave-free NV sensing protocols could be deployed in this case[23]. In addition, a microwave-free sensing protocol and a diamond magnetometer have been used to create conductivity maps of conductive objects with sub-mm spatial resolution[24] in a substantial background field up to 100 mT. This procedure could be easily adapted to magnetic susceptibility measurements as well. Fluxgate sensors also offer ease of use and low cost but have lower sensitivity and poor spatial resolution. For sensors with a dynamic range that can accommodate environmental noise, it would be possible to lower the magnetic shielding demands by, for example, compensating this noise using an array of background sensors.

The induced field amplitude is proportional to the field produced by the solenoid, therefore the measurement of the magnetic susceptibility distribution could be made more sensitive by increasing the solenoid field. In addition, varying or modulating the magnetic field of the solenoid would also provide further means of separating the effects of magnetic susceptibility and internal currents.

## Conclusion

We have demonstrated the ability to detect states of charge and defects via magnetic susceptibility measurements and measurements of magnetic fields produced by internal currents using atomic magnetometers. Magnetic shielding and a solenoid were arranged in such a way that the sensors do not directly measure the magnetic field produced by the solenoid, while retaining full sensitivity to the fields originating from and induced within the battery. Battery-cell defects could be detected via fast and sensitive recordings of localized magnetic-susceptibility and internal-current-induced changes. The apparatus is in principle scalable and could be adapted to measure large-format cells as well, as used, for example, in electric vehicles. This approach could pave the way to a low-cost high-throughput diagnostic device for battery assessment and development.

## Methods

**Battery cells**
Measurements were performed on Li-ion pouch cells (Powerstream PGEBNMU53040) with dimensions of approximately 3x4x0.5 cm$^3$. Energy-dispersive X-ray spectroscopy of the cathode materials revealed the composition as 44.25% Co, 33.20% O, 3.11% Mn, 13.95% C , 5.04% Ni, 0.1% Ti and 0.24% P in weight percentage. Supplementary Figures S1-S2 and Table S1 summarize the detailed measurement results.

The magnetic field around each battery with the solenoid off was measured to be less than 20 pT in any direction before performing experiments. This field (due to remnant magnetization) was mapped for each cell with the solenoid field off and subtracted from the induced field maps.

**Solenoid, magnetic shield, magnetometers, and data acquisition**
All measurements were performed inside a Twinleaf MS-2 magnetic shield, operated with end-caps removed. The magnetic shield had four cylindrical layers of mu-metal shielding material with the innermost shield diameter of 180 mm and the outermost shield diameter of 304.5 mm. The overall length of the outermost shield was 620 mm. At the center of the shield cylinders, the laboratory magnetic fields were reduced by a factor of $10^4$ in the direction parallel to the cylinder axis, and $10^5$ in the transverse direction.
With end-caps removed and compensation fields applied, the magnetic-field on top and below the center of the solenoid within the magnetic shield region was measured to be below 10 pT. This value was well within acceptable background levels for the susceptibility measurements. Acquiring measurements within an open shield reduces baseline noise in the magnetometer to environmental noise dominated ~40 pT /√Hz in the *x*-direction and ~200 pT /√Hz in the *z*-direction (larger due to open cylinder arrangement), within the sensor bandwidth of DC to 100 Hz. The measurement can be susceptible to large transient magnetic artifacts and long-term drifts resulting from other equipment in the laboratory or in the building, therefore the background fields were monitored using a fluxgate magnetometer to ensure that measurements were not taken during large background-field fluctuations.

The coil was fabricated with 3560 turns of 0.56 mm diameter wire and uniform spacing on a 1 m long aluminum mount. Figure S3 shows the cross section of the flat solenoid design as well as a picture of the solenoid itself. A current of 3 mA was fed through the coil to produce a field of $2 \cdot 10^{-5}$ T at the center of the coil, measured using a small-sized fluxgate sensor. Two QuSpin zero-field rubidium-vapor spin-exchange relaxation free (SERF) magnetometers[17] (Generation 1) were used[22]. Each sensor was placed 2 cm above or below the center of the solenoid. Only the data from a single sensor is required for all the measurements shown in this work, and the second sensor was included for redundancy.

To measure the discharge behavior of the cell, it was discharged by drawing 2-3 mA of current over a 1 kΩ resistor for 30 min at a time. The cell was subsequently disconnected from the resistor for a further 30 min. During this rest period, the magnetic field at a single point above the battery was measured continuously. The position of the sensor for these measurements was determined by choosing a maximum point in the magnetic field map (upper left corner of the battery in Figure 1c). An analog input channel of a DAQ was connected to the circuit during both discharging and rest conditions. This discharge and rest cycle were repeated over 24 hours. During this periodic discharge measurement, the solenoid field was on, so that susceptibility changes could be monitored. The same measurement performed with the solenoid field off yielded no induced field in the cell, and the relaxation time constants for internal currents were determined to be similar as when the solenoid field was turned on. The discharge capacity of the battery at each point was determined by taking the average current measured in the discharge circuit during a discharge period, multiplied by the discharge time.

The cell was transported through the solenoid region and past the sensors using a conveyor belt made of vinyl tape, chosen for its low-friction backing and its non-magnetic properties. The belt was moved via servo motors that were placed outside the shielded region. Each servo motor was mounted on a translation stage (Thorlabs KMTS50E/M). This arrangement allowed the entire belt to move in the *x*-direction with 0.1 mm precision, enabling scanning measurements of the batteries. The sensors measured the *x* and *z* magnetic-field components in a plane parallel to the battery (see Figure 1).

The QuSpin electronics provide dedicated analog voltage output channels corresponding to the measured magnetic field in the *x*- or *z*- directions[17]. These voltages were measured using a National Instruments DAQ (NI-9205). The motion of the conveyor belt was controlled via a custom Python program that sequentially moved the belt forward by approximately 40 cm and was then reset to the original longitudinal position while the translation stages laterally moved the belt by 0.5 mm for the next scan. The data acquisition was triggered by a mechanical switch that was tripped at the end of the conveyor belt reset movement.

Each map took approximately 15 min to acquire, limited by the speed of the motors used, and was obtained using an average of six runs. The magnetic field data was filtered for line noise and harmonics, and smoothed using a 20-point window.

**Magnetic field inversion**
Regularized magnetic field inversion was performed using a dipolar field kernel and a truncated singular value decomposition[25]. The volume magnetic susceptibility distribution was obtained as an average value across the thickness of the cell ($y$-direction). Singular values of the dipolar kernel transformation matrix were truncated below a value of $5\times10^{-7}$ T to achieve regularization and stability of the inversion.

**Acknowledgements**
The work was funded in part by a grant by the U.S. National Science Foundation under award CBET 1804723 and the German Federal Ministry of Education and Research (BMBF) within the Quantumtechnologien program (FKZ 13N14439) and the DFG through the DIP program (FO 703/2-1).